# Hot Electron Dynamics in Zincblende and Wurtzite GaN

C. G. Rodrigues (a), V. N. Freire  (b), J. A. P. da Costa (c),
A. R. Vasconcellos (a), and R. Luzzi (a)

*(a) Instituto de Física Gleb Wataghin, Universidade Estadual de Campinas,
Caixa Postal 6165, 13083-970 Campinas, São Paulo, Brazil*

*(b) Departamento de Física, Universidade Federal do Ceará, Caixa Postal 6030,
Campus do Pici, 60455-760 Fortaleza, Ceará, Brazil*

*(c) Departamento de Física, Universidade Federal do Rio Grande do Norte,
Caixa Postal 1641, 59072-970 Natal, Rio Grande do Norte, Brazil*
`cloves@pucgoias.edu.br`

A theoretical investigation of the excess energy dissipation of highly excited photoinjected electrons in both wurtzite and zincblende GaN is presented. The calculations are performed by solving numerically coupled quantum transport equations for the carriers and the acoustic, transversal and longitudinal optical phonon in order to derive the evolution of their nonequilibrium temperatures, dubbed quasi-temperatures. It is shown that the electron energy dissipation is always faster in the wurtzite structure than in zincblende, both occurring in a subpicosecond time scale ($< 0.2$ ps).

## 1. Introduction

The technological applications of III–V nitride semiconductors are inducing considerable effort in the understanding of their basic properties [1 to 4]. In particular, a good deal of interest is centered on the research of GaN because of its use in the confinement layers of most nitride based quantum wells. Photoexcited GaN excess carrier energy relaxation is one important subject that is only recently receiving attention [5 to 9], as a result of the relevance for understanding how the excess carrier energy dissipation influences the working of devices. Information on the carrier dynamics in both wurtzite and zincblende GaN phases is of interest because the former presents many advantages related to the growth process, and the latter offers higher saturated electron drift velocity, easy cleavage, etc., what is of importance for device performance [10 to 12].

We here consider the carrier relaxation kinetics in a highly excited photoinjected plasma (HEPS) generated in both wurtzite and zincblende GaN. We recall that a HEPS consists of electrons and holes as mobile carriers (created by an intense ultrashort laser pulse), which are moving in the background of lattice vibrations. These photoexcited carriers with a concentration $n$ are initially narrowly distributed around the energy levels in conduction and valence bands separated by the photon energy $\hbar\Omega_L$. After this initial stage they proceed to redistribute their excess energy as a result of strong Coulomb interaction. At the same time relaxation processes occur transferring the excess energy to the lattice and also to the media surrounding the active volume of the sam-ple. On the other hand, the concentration diminishes in time via the processes of recombination and ambipolar diffusion [13].

Assuming valid the modelling used in [6], we take the nonequilibrium carrier distributions in photoexcited GaN as described by distribution functions characterized by quasi-temperatures well above that of the lattice. The energy relaxation of the carriers follows mainly through scattering processes with phonons (dominated by LO phonons) [5, 8]. We study the hot electron dynamics using coupled quantum transport equations for the carrier quasi-temperature (after Coulomb thermalization) and the acoustic, transversal and longitudinal optical phonon temperatures to have a picture of their excess energy dissipation.

## 2. Model and Method

We deal theoretically with the hot-electron dynamics in zincblende and wurtzite GaN, which is basically a problem involving a nonlinear nonequilibrium kinetic, resorting to a theory built within the framework of a particular nonequilibrium ensemble formalism, the so-called *Nonequilibrium Statistical Operator Method* (NESOM) [14, 15] and Zubarev's approach is used [16]. Moreover, we introduce the Markovian approximation to the theory [15, 17]. To proceed further it is necessary to characterize the nonequilibrium macroscopic (i.e. nonequilibrium thermodynamic [18]) state of the system, the so-called kinetic stage appropriate for its description in the given experimental conditions (as discussed in [19]) mainly considering the extension of the pulse of the exciting laser source and the resolution time of the detector. First, we notice that a single-quasiparticle description (allowing to introduce a band structure for the electron energies and the use of the random phase approximation) can follow after a very initial transient time of the order of a period of a plasma wave, which is typically in the tenfold femtosecond scale. The next kinetic stage corresponds to a description of the carriers in terms of distribution functions in single-particle band energy states. Next, under the action of Coulomb interaction and carrier–phonon collisions there follows an internal thermalization of the carriers whose nonequilibrium thermodynamic state can be characterized by the time-dependent density $n(t)$ and quasi-temperature $T^*(t)$ [19, 20]. In most cases a third kinetic stage is characterized by the mutual thermalization of carriers and optical phonons, followed by the attainment of final equilibrium with the acoustical phonons and the external thermal reservoir.

## 3. Results and Discussions

We present here a first partial study concerning the photoexcited carrier dynamics in zincblende and wurtzite GaN. We consider an ultrashort pumping laser pulse producing an initial concentration of carriers $n = 1.0 \times 10^{18}$ cm$^{-3}$, and excess energy of 1.2 eV to avoid intervalley scattering. We began to consider the second kinetic stage, i.e. we assume that Coulomb thermalization among the carriers has already occurred, and we take for them an initial quasi-temperature $T^*(0) = 4640$ K (corresponding to the excess energy per carrier of 1.2 eV). The temperature of the lattice is 300 K, and we consider the case a very good thermal contact keeps the optical phonons constant at the reservoir temperature. We derive the equations of evolution for the carrier quasi-temperature and concentration through the NESOM within the Markovian approximation. We omit the details of the calculations which follow the same scheme as those in [13] for GaAs. Table 1 presents the zincblende and wurtzite GaN parameters used in the calculations.

Table 1

| Parameter | GaN wurtzite | GaN zincblende |
|---|---|---|
| electron effective mass ($m_0$) | 0.19 [24] | 0.19 [24] |
| heavy-hole effective mass ($m_0$) | 2.0 [24] | 0.86 [24] |
| band gap energy (eV) | 3.5 [25] | 3.4 [26] |
| lattice parameter $a(c)$ (Å) | 3.189 (5.185) [27] | 4.5 [28] |
| LO-phonon energy (meV) | 92 [29] | 92 [29] |
| static dielectric constant $\varepsilon_0$ | 9.5 [27] | 9.5 [27] |
| optical dielectric constant $\varepsilon_\infty$ | 5.35 [27] | 5.35 [27] |
| mass density (g/cm$^3$) | 6.095 [30] | 6.095 [30] |
| acoustic deformation potential $E_{1e}$ (eV) | 8.3 [31] | 10.1 [32] |
| acoustic deformation potential $E_{1h}$ (eV) | 4.28 [33] | 4.02 [33] |
| piezoelectric constant (C/m$^2$) | 0.375 [34] | 0.560 [34] |

In Fig. 1 we show the evolution of the quasi-temperature for the times involved (subpicosecond scale) when the concentration $n(t)$ is practically unaltered. Inspection of Fig. 1 tells us that there occurs a very rapid relaxation of the carrier excess energy to the lattice (with a characteristic time of roughly 100 fs), together with a final thermalization and equilibrium in, say, 200 fs. This is a result of the effect of the intense Fröhlich interaction between carriers and LO-phonons in this strongly polar semiconductor. The energy dissipation is shown to be faster in wurtzite GaN than in zincblende GaN, but in both cases follows with a characteristic relaxation time in the hundredfold femtosecond scale.

## 4. Final Remarks

We return to the question of the description of the evolution of the nonequilibrium macrostate of the system in terms of a hierarchy of relaxation times defining successive kinetic stages. We recall that Fig. 1 was obtained starting at the third kinetic stage after

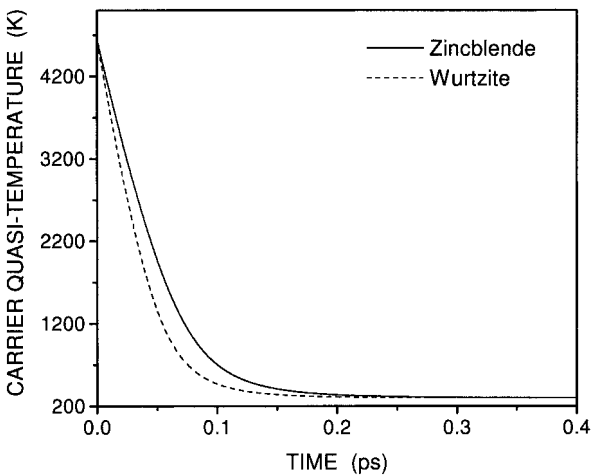

Fig. 1. Time evolution of the carrier quasi-temperature in wurtzite (dashed line) and zincblende (solid line) GaN. The initial photoinjected carrier density and temperature are $n = 1.0 \times 10^{18}$ cm$^{-3}$ and $T^*(0) = 4640$ K, respectively. The GaN lattice is maintained at 300 K

assuming Coulomb thermalization of the carriers. But then our results, indicating a relaxation time to the lattice in the hundredfold femtosecond scale requires that Coulomb thermalization should have followed in the tenfold femtosecond scale. But this is dubious. For other materials, namely III–V an IV compounds thermal relaxation follows in, roughly 500 fs to 1 ps ([21] and [22], respectively). Gross estimates for GaN give also values in a picosecond scale for the concentration we have used. Thus, in GaN (and also in all III–N compounds) the strong polar interaction (Fröhlich's carrier–phonon interaction) leads to relaxation times for the carrier excess energy out of equilibrium, comparable with the one resulting from internal Coulomb interaction.

Hence, our study gives the strong indication that for III–N compounds, the analysis of the ultrafast transient which shows the relaxation in the highly-excited photoinjected plasma requires to proceed to its description in the second kinetic stage. This means that one must include as basic macroscopic quantities the population of the carriers in the different band-energy states and to carry out the joint calculation of relaxation effects due to carrier–carrier interaction and carrier–phonons interaction. We are in the process of deriving the corresponding theory and calculations [23].